# Fission of Multiply Charged Alkali Clusters in Helium Droplets - Approaching the Rayleigh Limit

Michael Renzler,[a] Martina Harnisch,[a] Matthias Daxner,[a] Lorenz Kranabetter,[a] Martin Kuhn,[a] Paul Scheier,[a,c] Olof Echt[a,b,c]

[a] *Institut für Ionenphysik und Angewandte Physik, University of Innsbruck, Technikerstrasse 25, A-6020 Innsbruck, Austria*
[b] *Department of Physics, University of New Hampshire, Durham, New Hampshire 03824, United States*

[c] Corresponding Authors
P. Scheier. E-mail: paul.scheier@uibk.ac.at. Phone: +43 512 507 52660. Fax: +43 512 507 2922.
O. Echt. E-mail: olof.echt@unh.edu. Phone: +1 603 862 3548. Fax: +01 603 862 2998.

**Abstract**
Electron ionization of helium droplets doped with sodium, potassium or cesium results in doubly and, for cesium, triply charged cluster ions. The smallest observable doubly charged clusters are $Na_9^{2+}$, $K_{11}^{2+}$, and $Cs_9^{2+}$; they are a factor two to three smaller than reported previously. The size of sodium and potassium dications approaches the Rayleigh limit $n_{Ray}$ for which the fission barrier is calculated to vanish, i.e. their fissilities are close to 1. Cesium dications are even smaller than $n_{Ray}$, implying that their fissilities have been significantly overestimated. Triply charged cesium clusters as small as $Cs_{19}^{3+}$ are observed; they are a factor 2.6 smaller than previously reported. Mechanisms that may be responsible for enhanced formation of clusters with high fissilities are discussed.

## 1. Introduction

The instability of highly charged droplets is a phenomenon of broad relevance that occurs on a wide range of length scales. Lord Rayleigh applied a continuum model to determine the maximum charge of macroscopic water droplets beyond which spherical droplets become unstable with respect to spontaneous deformation. This so-called Rayleigh limit occurs when the Coulomb energy exceeds twice the surface energy.[1] Charge-separation of droplets near the Rayleigh-limit is of practical relevance to electrospray sources and inkjet printing[2,3] and it may prove useful to separate carbon nanotubes.[4] The disintegration of micron-sized liquid glycol droplets at the Rayleigh limit has been imaged by Leisner and coworkers.[5,6] Imaging by Mason et al. revealed that the explosive behavior of millimeter-sized alkali droplets injected into water is due to the Rayleigh instability.[7]

At the other end of the length scale one encounters fission of atomic nuclei.[8] In between, on the mesoscopic scale, charge-driven instabilities manifest themselves in the absence of multiply charged atomic clusters $A_n^{z+}$ in mass spectra below an experimental appearance size $n_{exp}$,[9] first documented by Sattler et al..[10] Numerous reports have explored the energetics and dynamics of cluster ions in the vicinity of $n_{exp}$ (for reviews, see [11-13]). Early work was directed at determining values of $n_{exp}$, fission barriers, and size distributions of fission fragments with a focus on van der Waals systems.[14-21] Later work focused on metal clusters and the role of electronic shell structure.[22-26]

Initially it was assumed that clusters of size $n_{exp}$ were near the Rayleigh limit $n_{Ray}$ at which the fission barrier vanishes. In other words, it was believed that their fissility was near $X = 1$ where $X$ is the ratio between the Coulomb energy and twice the surface energy.[1] A more useful relation, when dealing with atomic or molecular clusters of size $n$, is $X = n_{Ray}/n$.[12,27] However, the observation of metastable fission, on the time scale of microseconds, revealed that fission near $n_{exp}$ is thermally activated and competes with evaporation of monomers.[15, 18, 21, 28-30] In most experiments cluster ions are "boiling hot;" their vibrational temperatures are set by the heat of evaporation and are difficult to vary.[31] Thus, in spite of the expected temperature dependence, experimental values of $n_{exp}$ have been highly reproducible.[32]

Given the competition between fission and evaporation, a meaningful definition of a theoretical appearance size[9] $n_a$ is the size at which the rate of thermally activated fission equals the rate of monomer evaporation.[27, 28] For multiply charged alkali clusters computed $n_a$ values agree, indeed, closely with experimental appearance sizes $n_{exp}$[24, 25, 34] but they are much larger than values computed for the Rayleigh limit $n_{Ray}$. The fissility of the smallest multiply charged alkali cluster ions produced by photoionization ranges from $n_{Ray}/n_{exp}$ = 0.28 to 0.38.[12, 35] Huber and coworkers demonstrated that colder cluster ions can be produced in collisions with highly charged atomic ions.[40] They attained fissilities as large as $X$ = 0.87 for ten-fold charged sodium clusters but for low charge states the increase in fissility over photoionization experiments was rather modest. The authors conjectured that fissilities reached in their experiment were limited by the initial cluster temperature.

Here we show that alkali clusters with fissilities approaching 1 can be identified in mass spectra if the neutral precursors are embedded in superfluid helium nanodroplets. The smallest detected doubly charged sodium clusters contain 9 atoms, a factor three less than in previous experiments;[41] for cesium dications $n_{exp}$ is even lower than values computed for $n_{Ray}$.[24, 34] Furthermore, we observe triply charged cesium ions that are a factor 2.6 smaller than reported previously by Martin and coworkers.[42]

## 2. Experimental

Helium nanodroplets were produced by expanding helium (Messer, purity 99.9999 %) at a stagnation pressure of 20 bar through a 5 µm nozzle, cooled to 9.5 K (sodium), 9.5 K (potassium) or 9.1 ± 0.1 K (cesium) into vacuum. Droplets that form in the expansion contain an average number of $10^5$ to $10^6$ atoms; at a temperature of 0.37 K they are superfluid.[43] The resulting supersonic beam traversed a pick-up cell into which sodium, potassium or cesium (Sigma Aldrich, purity 99.95 % based on a trace metals analysis) were vaporized. The doped helium droplets were ionized by electrons at or below 70 eV; cations were accelerated and extracted by a pulsed voltage into a high-resolution reflectron-type time-of-flight mass spectrometer. The mass resolution (measured at full-width-at-half maximum) was 2900. Additional experimental details have been provided elsewhere.[44]

Mass spectra were evaluated by means of custom-designed software.[45] The routine includes automatic fitting of mass peaks and subtraction of background; it explicitly considers isotopic patterns of all ions. The abundance of specific ions (specific value of $n$) is derived by a matrix method. Critical regions were also inspected visually and fitted manually in order to ensure that impurity peaks and background were properly taken into account.

## 3. Results

Fig. 1 displays sections of mass spectra recorded for sodium- and cesium-doped helium droplets. The abscissa has been labeled in terms of the size-to-charge ratio $n/z$ of $A_n^{z+}$ (A = Na, Cs) or, in other words, the mass/charge ratio $m/z$ divided by the atomic mass of sodium or cesium, respectively (these elements are monisotopic).

Fig. 1a shows sections of a spectrum of sodium-doped droplets in the vicinity of $n/z$ values that are half-integer, i.e. sections where doubly charged ions containing an odd number of atoms would appear (even-numbered $Na_n^{2+}$ cannot be identified because they coincide with singly charged cluster ions twice as large). Each section covers a range of $\Delta n/z = \pm 0.01$, equivalent to $\Delta m/z = \pm 0.23$ u. The spectrum clearly establishes the occurrence of odd-numbered $Na_n^{2+}$ for $n \geq 9$. The abundance of $Na_5^{2+}$ and $Na_7^{2+}$ (marked by vertical dashed lines) is an order of magnitude weaker than that of $Na_9^{2+}$, and statistically not significant. The smallest dication observed by Martin and coworkers in a multi-step photoionization study using an F$_2$ laser (photon energy 7.9 eV) was Na$_{27}^{2+}$.[41, 42]

Fig. 1b shows sections of a spectrum that reveal the appearance of Cs$_n^{2+}$ for $n \geq 9$. Each section covers a range of $\Delta n/z = \pm 0.01$, equivalent to $\Delta m/z = \pm 1.3$ u. The ion yield at the position of Cs$_7^{2+}$ (marked by a vertical dashed line) is not statistically significant. The smallest dication observed by Martin and coworkers was Cs$_{19}^{2+}$.[42]

Fig. 1c shows sections where Cs$_n^{3+}$ ions would appear ($3n/z$ = integer). Solid lines represent gaussians fitted to the data. Their positions were fixed at integer values of $3n/z$ and their fixed widths computed from the mass resolution of the instrument. The right-most trace shows Cs$_{49}^{3+}$, the smallest trication identified by Martin and coworkers.[42] Our spectra reveal trications as small as Cs$_{19}^{3+}$. However, their abundance is two orders of magnitude less than that of small Cs$_n^{2+}$ seen in Fig. 1b. Establishing the existence or absence of even smaller trications is challenging though. Cs$_{17}^{3+}$ appears as a small peak whose statistical significance is doubtful. Cs$_{16}^{3+}$ is masked by a strong impurity peak. Cs$_{14}^{3+}$ would occur at $\Delta m/z = 0.178$ u to the left of He$_{155}^+$ as indicated by a vertical dashed line in Fig. 1c. A fit of two gaussians in this region (solid line) shows no significant contribution from Cs$_{14}^{3+}$.[46]

We now turn to potassium which has two isotopes of significant abundance, $^{39}$K (mass 38.964 u, natural abundance 93.26 %) and $^{41}$K (40.962 u, 6.73 %). Combined with the presence of ions that result from water impurities this leads to congested spectra. Fig. 2 shows sections of a mass spectrum that reveal the presence of odd-numbered K$_n^{2+}$ for $n \geq 11$. Each panel covers the four most abundant isotopologues containing zero to three $^{41}$K; the expected positions of these isotopologues are marked by symbols.

In Fig. 2d all four K$_{19}^{2+}$ isotopologues are resolved. A set of four gaussians with individually chosen backgrounds was fitted to the four peaks. Positions and relative peak heights were fixed at the theoretical values; a common value was chosen for the width. The fit (solid cyan line) matches the experimental data very well. All other significant mass peaks in Fig. 2d are identified in the Figure or its Caption. Spectra covering smaller odd-numbered K$_n^{2+}$ were analyzed similarly. Isotopologues overwhelmed by other peaks (He$_m^+$, He$_m$K$^+$ or reaction products of H$_2$O with K$_n^+$) were excluded from the fit. For K$_{13}^{2+}$ and K$_{15}^{2+}$ three isotopologues are clearly seen in Fig. 2. The yield of K$_{11}^{2+}$ is weak but the amplitudes of the first two isotopologues, containing zero or one $^{41}$K, are non-zero with a statistical significance greater than 4σ. K$_9^{2+}$ and K$_7^{2+}$ could not be identified.

Even-numbered doubly charged potassium clusters can be identified for $n \geq 20$ thanks to the presence of isotopologues. Other doubly charged cluster ions that could be identified are H$_2$OK$_n^{2+}$ ($n \geq 16$), OHK$_n^{2+}$ and HK$_n^{2+}$.

Fig. 3 summarizes our results for doubly charged cluster ions. Solid dots represent their abundances, deduced from mass spectra after correction for background, other interfering mass peaks, and, for potassium, the presence of isotopologues. Odd-numbered Na$_n^{2+}$ and Cs$_n^{2+}$ are observed with high statistical significance for $n \geq 9$; the yields of Na$_5^{2+}$, Na$_7^{2+}$ and Cs$_7^{2+}$ are zero within the error bars. Diamonds represent the results of previous experiments by Martin and coworkers who identified Na$_n^{2+}$ for $n \geq 27$ and Cs$_n^{2+}$ for $n \geq 19$.[41, 42] Their data and error bars were estimated from published mass spectra and scaled to match our values at $n = 27$ (Na) and 21 (Cs).

The results for potassium-doped helium droplets are summarized in Fig. 3c. Odd-numbered K$_n^{2+}$ ions appear for $n \geq 11$, even-numbered dications can be identified for $n \geq 20$. The smallest potassium dication reported previously, in a multiphoton ionization study by Brechignac and coworkers using a nitrogen laser (photon energy 3.67 eV), was K$_{19}^{2+}$.[23] Their data (estimated from a mass spectrum) are shown as diamonds. Their mass resolution was not sufficient for the identification of even-numbered K$_n^{2+}$.

Our results for Cs$_n^{3+}$ are compiled in Fig. 4. The experimental appearance size is $n_{\text{exp}} = 19$ or, perhaps, 17. The abundances of the next smaller trications (C$_{14}^{3+}$, C$_{13}^{3+}$, and C$_{11}^{3+}$) that do not strongly interfere with other ions are all zero within the statistical error. $Cs_{20}^{3+}$ forms a remarkably strong peak. All larger trications with non-integer values of $n/3$ are statistically significant except perhaps $Cs_{26}^{3+}$. Diamonds represent the results of previous experiments by Martin and coworkers who identified trications for $n \geq 49$.[42] Their data were estimated from a mass spectrum and scaled. The difference in the size dependence of the two data sets beyond $n \approx 55$ is likely caused by differences in the size distributions of neutral precursors and the way ions are formed; the difference is not relevant for the present discussion.

Mass spectra of helium droplets doped with sodium and other alkalis produce copious amounts of alkali cations (monomers as well as dimers) complexed with helium.[47-49] These ions are often referred to as snowballs because electrostrictive forces solidify the helium matrix that surrounds the ion. One may wonder if $He_mA_n^{2+}$ (A = sodium, potassium, cesium) dications can be detected for $n > 1$. The answer is no, but that is hardly surprising. Consider, for example, sodium. The most abundant dication is $Na_{27}^{2+}$. Among all possible $He_mNa_{27}^{2+}$ ions the detection threshold would be best for $He_2Na_{27}^{2+}$; the ion could be identified if its abundance were at least 5 % that of $Na_{27}^{2+}$. To put this in perspective: the abundance of $Na_n^+$ with one or two helium attached is about 10 % for $Na^+$, 4 % for $Na_2^+$, but only 0.5 % for $Na_3^+$ (relative to the corresponding bare alkali ions). We cannot safely identify any singly charged $He_mNa_n^+$ ions containing more than 3 sodium atoms. Furthermore, although we observe atomic $Na^{2+}$ its abundance is low; no $He_mNa^{2+}$ ions are detected. However, the abundance of $Cs^{2+}$ is much higher and $He_mCs^{2+}$ ions appear at the one-percent level. The only other complexes of dications with helium reported so far (by the Meiwes-Broer group using femtosecond laser pulses[50] [51]) are $Mg^{2+}$, $Ag^{2+}$ and $Pb^{2+}$.

**4. Discussion**

The main results of our experiments are
1) Detection of $Na_n^{2+}$, $K_n^{2+}$, $Cs_n^{2+}$, and $Cs_n^{3+}$ that are much smaller (by a factor 3 for $Na_n^{2+}$) than previously observed,
2) Close agreement of the experimental appearance size $n_{exp}$, the size of the smallest observed multiply charged clusters, with the Rayleigh limit $n_{Ray}$, the size for which the calculated fission barrier vanishes (i.e. cluster ions with fissility $X = 1$).

We emphasize that observation (1) is not merely the result of enhanced detection efficiency or resolution. These factors may lower $n_{exp}$ but hardly by more than a few percent.[32] Instead, in the present work the dramatic decrease of $n_{exp}$ results from enhanced production of small multiply charged ions. This is apparent from the distributions displayed in Figures 3 and 4. For example, in our work the abundance ratio of $Cs_{19}^{2+} : Cs_{21}^{2+}$ is 2.4 times larger than reported by Martin and coworkers.[42] For $K_{19}^{2+} : K_{21}^{2+}$ our value is five times larger than the value reported by Brechignac et al..[23] For $Na_{25}^{2+} : Na_{27}^{2+}$ we obtain about 1 : 5; Martin and coworkers[52] could not observe $Na_{25}^{2+}$ even though the mass resolution of their instrument (20 000) was seven times better than that of our instrument. From their spectrum [41] we estimate that their ratio was below 1 : 500.

Before discussing possible mechanisms for the enhanced production of small multiply charged cluster ions we turn to observation (2). Values of $n_{Ray}$ and appearance sizes $n_a$ calculated for doubly charged sodium, potassium and cesium clusters and for Cs trications are listed in Table 1 together with experimental appearance sizes $n_{exp}$ observed in past experiments and in the present work. One notes that i) previously reported $n_{exp}$ values are close to $n_a$ (about 20 % lower for K and Cs, 10 % higher for Na), ii) our current $n_{exp}$ values are close to $n_{Ray}$. In other words, the fissility of the smallest observed multiply charged clusters is close to 1.

As $n_{Ray}$ marks the size where the fission barrier vanishes one should always find $n_{Ray} < n_{exp}$. The values $n_{Ray} = 6$ or 7 calculated for $Na_n^{2+}$ by Barnett et al.,[53] Blaise et al.[54] and Vierira et al.[25] are consistent with this inequality, but the values 11 or 12 computed by Li et al.[24] and Garcias et al.[34] are not. Similarly, $n_{Ray}$ values calculated for $K_n^{2+}$ by Brechignac et al.[36] and Vieira et al.[25] are compatible with our experimental value for $n_{exp}$ while other calculated values are too high. In general, $n_{Ray}$ values obtained by ab-initio molecular dynamics simulations of $K_n^{2+}$ [36] and $Na_n^{2+}$ [53, 54] tend to be smaller, more compatible with our experimental results. For $Cs_n^{2+}$ all calculated $n_{Ray}$ values exceed our value $n_{exp} = 9$. For $Cs_n^{3+}$ we observe $n_{exp} = 19$ (or perhaps 17) which is in reasonable agreement with the Rayleigh limit $n_{Ray} = 19$ deduced using a many-body potential based on local density calculations.[24] Another value, $n_{Ray} = 27$ calculated by applying a semi-empirical model to compute fission barriers[34] is clearly too large.

Why is our experimental approach so efficient in the production of long-lived multiply charged ions near $X = 1$? There are two major differences between the present and previous studies, i) the mechanism by which ions are formed, and ii) the presence of the helium droplet that cools the neutral precursor to 0.37 K and may help to cool and/or cage the cluster after ionization.[56]

Electron ionization of doped helium droplets involves one or more intermediate steps. If the energy of the incident electrons exceeds the ionization threshold of helium (24.59 eV), ionization proceeds primarily via formation of $He^+$ which will hop, on the time scale of femtoseconds, by resonant charge transfer.[57] $He^+$ may ionize the dopant, or localize by forming a vibrationally excited $He_2^+$. We have measured the yield of $Cs_n^{z+}$ ($z = 1, 2$) versus electron energy and found a stepwise onset near 21 eV. This implies that ionization proceeds by Penning ionization, involving metastable, electronically excited $He^*$, $He_2^*$, or anions.[58, 59] Moreover, neutral cesium clusters will reside on the surface of the droplet.[60] Neutral sodium and potassium clusters will reside on the surface if they contain fewer than ≈20 or ≈80 atoms, respectively.[44, 60, 61] Location on the surface favors Penning ionization over ionization by $He^+$.[61]

According to Huber and coworkers, the ability to form multiply charged clusters near $X = 1$ is mainly limited by the initial cluster temperature.[40] Our experiments involve ultracold (0.37 K) precursors but it is questionable if this is sufficient to suppress fission of nascent multiply charged clusters given that our mass spectra indicate strong fragmentation: In Fig. 3c $K_{22}^{2+}$ stands out particularly strong. According to the spherical jellium model this ion, containing 20 valence electrons (in the configuration $1s^2 1p^6 1d^{10} 2s^2$) is particularly stable.[62] As growth of neutral clusters in helium droplets is a statistical process, the enhanced $K_{22}^{2+}$ abundance must be a result of ionization induced fragmentation.[63]

Thus, we face a conundrum: evidence for fragmentation (appearance of magic numbers) and, at the same time, evidence for suppression of fission ($X$ approaching 1). We will sketch two factors that may be relevant but more work is needed to pin down the exact mechanism. First, it is conceivable that multiply charged cluster ions form by successive ionization, involving two or more incident electrons as observed in some of our past work.[59, 64] Penning ionization at the surface will result in an alkali cluster ion $A_n^+$ which will dive into the droplet and be cooled to 0.37 K. At some later time (within microseconds) another incident electron will form $He^*$ or $He^{*-}$ which will form $A_m^{2+}$ ($m \leq n$) by Penning ionization. Although successive ionization of bare clusters by photons does not result in multiply charged clusters of high fissility,[42] effective cooling and caging of singly charged cluster ions submersed in helium may be sufficient to suppress fission. Unfortunately the low ion yield of the newly observed small multiply charged cluster ions makes a systematic study of the dependence of their yield on the electron emission current all but impossible.

Second, fission proceeds on a slower time scale than monomer evaporation because the frequency of collective modes is relatively low; viscosity will further slow down the reaction.[14, 27, 65] The extension of the time scale may be particularly strong for solid precursors.[12] Möller and coworkers have reported a dramatic difference in the fission pattern of solid xenon clusters as compared to liquid neon clusters for fissilities near $X = 1$.[37] The slower the rate of fission the larger the chance that the helium matrix can remove the excess energy and quench the reaction, provided the fissility is below 1.

The search for long-lived multiply charged clusters with high fissilities is motivated by the expectation of new physics.[12, 37, 40, 55] For metal clusters one expects that fission will change from highly asymmetric (emission of trimer ions) to symmetric as $X$ approaches 1.[12, 26, 66] Moreover, the strong effects of electronic shell structure that govern fission of warm, liquid clusters with $X$ well below 1 may be greatly reduced or even absent.[12] Fission of doubly charged alkali clusters is often likened to alpha decay of radioactive nuclei because of the preference for emission of trimer ions which are favored by electronic shell effects.[27] However, with increasing fissility and decreasing temperature the balance shifts toward more symmetric fission;[26, 66] this shift will also contribute to a slowing of the fission reaction.

## 5. Conclusions

Electron ionization of alkali clusters embedded in helium droplets results in doubly and triply charged clusters that are much smaller than observed previously; their fissility is approaching $X = 1$. We have discussed some factors that might enhance the yield of these small multiply charged cluster ions by suppressing rapid fission. Preparing clusters of high fissility is a first step; future experiments will have to characterize their fission reactions, for example by gentle heating with low-energy photons. Molecular dynamics simulations of multiply charged clusters embedded in helium would be helpful to unravel details.

**Acknowledgements**

This work was supported by the Austrian Science Fund, Wien (FWF Projects I978 and P26635).

**References**
1. L. Rayleigh, *Phil. Mag.*, 1882, **14**, 184-186.
2. W. Gu, P. E. Heil, H. Choi and K. Kimb, *Appl. Phys. Lett.*, 2007, **91**, 064104.
3. S. Consta and A. Malevanets, *Phys. Rev. Lett.*, 2012, **109**, 148301.
4. G. T. Liu, Y. C. Zhao, K. H. Zheng, Z. Liu, W. J. Ma, Y. Ren, S. S. Xie and L. F. Sun, *Nano Lett.*, 2009, **9**, 239-244.
5. D. Duft, T. Achtzehn, R. Müller, B. A. Huber and T. Leisner, *Nature*, 2003, **421**, 128-128.
6. E. Giglio, B. Gervais, J. Rangama, B. Manil, B. A. Huber, D. Duft, R. Müller, T. Leisner and C. Guet, *Phys. Rev. E*, 2008, **77**, 036319.
7. P. E. Mason, F. Uhlig, V. Vanek, T. Buttersack, S. Bauerecker and P. Jungwirth, *Nature Chem.*, 2015, **7**, 250-254.
8. N. Bohr and J. A. Wheeler, *Phys. Rev. A*, 1939, **56**, 426-450.
9. In the literature *appearance* sizes are often called *critical* sizes. The terms may pertain to experimental as well as theoretical values; we refer to experimental values as $n_{exp}$ and theoretical values as $n_a$, respectively. The Rayleigh limit $n_{Ray}$ is the size at which the fissility $X$ equaks 1, i.e. the calculated fission barrier vanishes.
10. K. Sattler, J. Mühlbach, O. Echt, P. Pfau and E. Recknagel, *Phys. Rev. Lett.*, 1981, **47**, 160-163.
11. O. Echt and T. D. Märk, in *Clusters of Atoms and Molecules II*, ed. H. Haberland, Springer-Verlag, Berlin, 1 edn., 1994, vol. 56, pp. 183-220.
12. U. Näher, S. Bjørnholm, S. Frauendorf, F. Garcias and C. Guet, *Phys. Rep.*, 1997, **285**, 245-320.
13. O. Echt, P. Scheier and T. D. Märk, *Comptes Rendus Physique*, 2002, **3**, 353-364.
14. J. G. Gay and B. J. Berne, *Phys. Rev. Lett.*, 1982, **49**, 194-198.
15. D. Kreisle, O. Echt, M. Knapp, E. Recknagel, K. Leiter, T. D. Märk, J. J. Saenz and J. M. Soler, *Phys. Rev. Lett.*, 1986, **56**, 1551-1554.
16. D. Kreisle, K. Leiter, O. Echt and T. D. Märk, *Z. Phys. D*, 1986, **3**, 319-322.
17. K. Leiter, D. Kreisle, O. Echt and T. D. Märk, *J. Phys. Chem.*, 1987, **91**, 2583-2586.

18. O. Echt, D. Kreisle, E. Recknagel, J. J. Saenz, R. Casero and J. M. Soler, *Phys. Rev. A*, 1988, **38**, 3236-3248.
19. N. G. Gotts, P. G. Lethbridge and A. J. Stace, *J. Chem. Phys.*, 1992, **96**, 408-421.
20. I. Mähr, F. Zappa, S. Denifl, D. Kubala, O. Echt, T. D. Märk and P. Scheier, *Phys. Rev. Lett.*, 2007, **98**, 023401.
21. G. H. Wu, X. J. Chen, A. J. Stace and P. Linse, *J. Chem. Phys.*, 2011, **134**, 031103.
22. M. Nakamura, Y. Ishii, A. Tamura and S. Sugano, *Phys. Rev. A*, 1990, **42**, 2267-2278.
23. C. Brechignac, P. Cahuzac, F. Carlier, J. Leygnier and A. Sarfati, *Phys. Rev. B*, 1991, **44**, 11386-11393.
24. Y. B. Li, E. Blaisten-Barojas and D. A. Papaconstantopoulos, *Phys. Rev. B*, 1998, **57**, 15519-15532.
25. A. Vieira and C. Fiolhais, *Phys. Rev. B*, 1998, **57**, 7352-7359.
26. C. Yannouleas, U. Landman, C. Brechignac, P. Cahuzac, B. Concina and J. Leygnier, *Phys. Rev. Lett.*, 2002, **89**, 173403.
27. P. Fröbrich, *Phys. Rev. B*, 1997, **56**, 6450-6453.
28. W. A. Saunders, *Phys. Rev. Lett.*, 1990, **64**, 3046-3049.
29. C. Brechignac, P. Cahuzac, F. Carlier and M. De Frutos, *Phys. Rev. Lett.*, 1990, **64**, 2893-2896.
30. S. Krückeberg, G. Dietrich, K. Lutzenkirchen, L. Schweikhard, C. Walther and J. Ziegler, *Z. Phys. D*, 1997, **40**, 341-344.
31. C. E. Klots, *Nature*, 1987, **327**, 222-223.
32. Experiments performed with high sensitivity may identify slightly smaller values of $n_{exp}$, see e.g. [33] and references therein.
33. M. Daxner, S. Denifl, P. Scheier and O. Echt, *Int. J. Mass Spectrom.*, 2014, **365–366**, 200–205.
34. F. Garcias, R. J. Lombard, M. Barranco, J. A. Alonso and J. M. Lopez, *Z. Phys. D*, 1995, **33**, 301-305.
35. It is possible to exceed $X = 1$, for example by photoionizing small singly charged clusters,[36] photoexciting core electrons in neutral cluster,[37] or exposing clusters to intense laser fields[38,39] but only the fragments of those clusters are observable.
36. C. Brechignac, P. Cahuzac, F. Carlier, M. De Frutos, R. N. Barnett and U. Landman, *Phys. Rev. Lett.*, 1994, **72**, 1636-1639.
37. M. Hoener, C. Bostedt, S. Schorb, H. Thomas, L. Foucar, O. Jagutzki, H. Schmidt-Böcking, R. Dörner and T. Möller, *Phys. Rev. A*, 2008, **78**, 021201.
38. A. E. Kaplan, B. Y. Dubetsky and P. L. Shkolnikov, *Phys. Rev. Lett.*, 2003, **91**, 143401.
39. A. Heidenreich, J. Jortner and I. Last, *Proc. Natl. Acad. Sci. USA*, 2006, **103**, 10589-10593.
40. F. Chandezon, S. Tomita, D. Cormier, P. Grübling, C. Guet, H. Lebius, A. Pesnelle and B. A. Huber, *Phys. Rev. Lett.*, 2001, **87**, 153402.
41. U. Näher, H. Göhlich, T. Lange and T. P. Martin, *Phys. Rev. Lett.*, 1992, **68**, 3416.
42. U. Näher, S. Frank, N. Malinowski, U. Zimmermann and T. P. Martin, *Z. Phys. D*, 1994, **31**, 191-197.
43. J. Tiggesbäumker and F. Stienkemeier, *Phys. Chem. Chem. Phys.*, 2007, **9**, 4748-4770.
44. L. An der Lan, P. Bartl, C. Leidlmair, H. Schöbel, R. Jochum, S. Denifl, T. D. Märk, A. M. Ellis and P. Scheier, *J. Chem. Phys.*, 2011, **135**, 044309.
45. S. Ralser, J. Postler, M. Harnisch, A. M. Ellis and P. Scheier, *Int. J. Mass Spectrom.*, 2015, **379**, 194-199.
46. The next close encounter between $Cs_n^{3+}$ and helium cluster ions occurs at $Cs_{29}^{3+}$ which would be separated from $He_{321}^+$ by $\Delta m/z = 0.083$ u. This coincidence causes no problem because $He_m^+$ ions are not discernible beyond $n \approx 100$.
47. S. Müller, M. Mudrich and F. Stienkemeier, *J. Chem. Phys.*, 2009, **131**, 044319.

48. M. Theisen, F. Lackner and W. E. Ernst, *J. Chem. Phys.*, 2011, **135**, 074306.
49. L. An der Lan, P. Bartl, C. Leidlmair, R. Jochum, S. Denifl, O. Echt and P. Scheier, *Chem. Eur. J.*, 2012, **18**, 4411-4418.
50. T. Döppner, T. Diederich, S. Gode, A. Przystawik, J. Tiggesbäumker and K. H. Meiwes-Broer, *J. Chem. Phys.*, 2007, **126**, 244513.
51. T. Döppner, T. Diederich, A. Przystawik, N. X. Truong, T. Fennel, J. Tiggesbäumker and K. H. Meiwes-Broer, *Phys. Chem. Chem. Phys.*, 2007, **9**, 4639-4652.
52. T. P. Martin, T. Bergmann, H. Göhlich and T. Lange, *J. Phys. Chem.*, 1991, **95**, 6421-6429.
53. R. N. Barnett, U. Landman and G. Rajagopal, *Phys. Rev. Lett.*, 1991, **67**, 3058-3061.
54. P. Blaise, S. A. Blundell, C. Guet and R. R. Zope, *Phys. Rev. Lett.*, 2001, **87**, 063401 -063404.
55. F. Chandezon, C. Guet, B. A. Huber, D. Jalabert, M. Maurel, E. Monnand, C. Ristori and J. C. Rocco, *Phys. Rev. Lett.*, 1995, **74**, 3784-3787.
56. A. Kautsch, M. Koch and W. E. Ernst, *Phys. Chem. Chem. Phys.*, 2015, **17**, 12310-12316.
57. A. M. Ellis and S. F. Yang, *Phys. Rev. A*, 2007, **76**, 032714.
58. A. Mauracher, M. Daxner, J. Postler, S. E. Huber, S. Denifl, P. Scheier and J. P. Toennies, *J. Phys. Chem. Lett.*, 2014, **5**, 2444–2449.
59. M. Renzler, M. Daxner, N. Weinberger, S. Denifl, P. Scheier and O. Echt, *Phys.Chem.Chem.Phys.*, 2014, **16**, 22466-22470.
60. C. Stark and V. V. Kresin, *Phys. Rev. B*, 2010, **81**, 085401.
61. L. An der Lan, P. Bartl, C. Leidlmair, H. Schöbel, S. Denifl, T. D. Märk, A. M. Ellis and P. Scheier, *Phys. Rev. B*, 2012, **85**, 115414.
62. W. A. De Heer, *Rev. Mod. Phys.*, 1993, **65**, 611-676.
63. Even-numbered $Na_n^{2+}$ and $Cs_n^{2+}$ cannot be identified because the elements are monisotopic. Triply charged cesium cluster ions would have closed shells if *n* = 23, 43, and 61. Those ions are identified in Fig. 4 but they do not stand out as particularly strong. $Cs_n^+$ would have closed electronic shells at *n* = 9 and 21, but only the first feature is reflected in the mass spectrum.
64. H. Schöbel, P. Bartl, C. Leidlmair, M. Daxner, S. Zöttl, S. Denifl, T. D. Märk, P. Scheier, D. Spångberg, A. Mauracher and D. K. Bohme, *Phys. Rev. Lett.*, 2010, **105**, 243402.
65. F. Calvo, *Phys. Rev. A*, 2006, **74**, 043202.
66. C. Brechignac, P. Cahuzac, B. Concina and J. Leygnier, *Phys. Rev. Lett.*, 2004, **92**, 083401.

**Table 1**
Lower size limits of multiply charged alkali clusters. $n_{Ray}$ denotes calculated Rayleigh limits where the fission barrier vanishes (fissility $X = 1$); $n_a$ are theoretical appearance sizes for which thermally activated fission rates equal evaporation rates. $n_{exp}$ specifies the smallest size observed in past experiments (publ) and the present work.

|  | $n_{Ray}$ (theory) | $n_a$ (theory) | $n_{exp}$ (publ) | $n_{exp}$ (this work) |
|---|---|---|---|---|
| $Na_n^{2+}$ | $6^{a,b}$, $7^c$, $11^d$, $12^e$ | $24^e$, $27^{c,d}$ | $27^{f,g}$ | 9 |
| $K_n^{2+}$ | $7^{c,h}$, $11^e$, $12^d$ | $22^c$, $24^e$ | $19^i$ | 11 |
| $Cs_n^{2+}$ | $11^e$, $12^d$ | $23^{d,e}$ | $19^g$ | 9 |
| $Cs_n^{3+}$ | $19^d$, $27^e$ | $51^d$, $57^e$ | $49^g$ | 19 (or 17) |

a  Ref. 53
b  Ref. 54
c  Ref. 25
d  Ref. 24
e  Ref. 34
f  Ref. 55
g  Ref. 42
h  Ref 36
i  Ref. 23

**Figure Captions**

Fig. 1
Sections of mass spectra showing the appearance of $Na_n^2$, $Cs_n^{2+}$, and $Cs_n^{3+}$ (panels a, b, c, respectively). Each section covers a range of ± 0.01 around the expected position of multiply charged ions. The smallest cluster ions that are deemed significant are $Na_9^{2+}$, $Cs_9^{2+}$, and $Cs_{19}^{3+}$; the expected positions of some smaller ions are marked by dashed lines.

Fig. 2
Sections of mass spectra showing the presence of $K_n^{2+}$ for $n$ = 11, 13, 15, 19 (panels a to d). The expected positions of the first four isotopologues (containing $p$ = zero to three $^{41}K$ isotopes) are marked by full symbols. Solid lines represent the results of fitting sets of three or four gaussians; positions and relative amplitudes were fixed at the expected values. Mass peaks labeled A, B or C correspond to $H_2O^{39}K_{n-1}{}^{41}K^+$, $OH^{39}K_n^+$, and $H_2O^{39}K_{n-2}{}^{41}K_2^+$, respectively.

Fig. 3
Full dots: Ion abundance of doubly charged $Na_n^{2+}$, $Cs_n^{2+}$ and $K_n^{2+}$ *versus* size $n$. Weak signals for small $n$ are re-plotted with enhanced amplitude. Solid lines connect data available for even and odd $n$; dashed lines connect data available for odd $n$ only. Diamonds represent results of photoionization studies by Näher et al.[41, 42] and Brechignac et al..[23] Their values were deduced from published mass spectra and scaled to match our values at $n$ = 27 (Na) or 21 (Cs, K).

Fig. 4
Full dots: Ion abundance of $Cs_n^{3+}$ *versus* size $n$. Diamonds represent results of photoionization studies by Näher et al.[42]

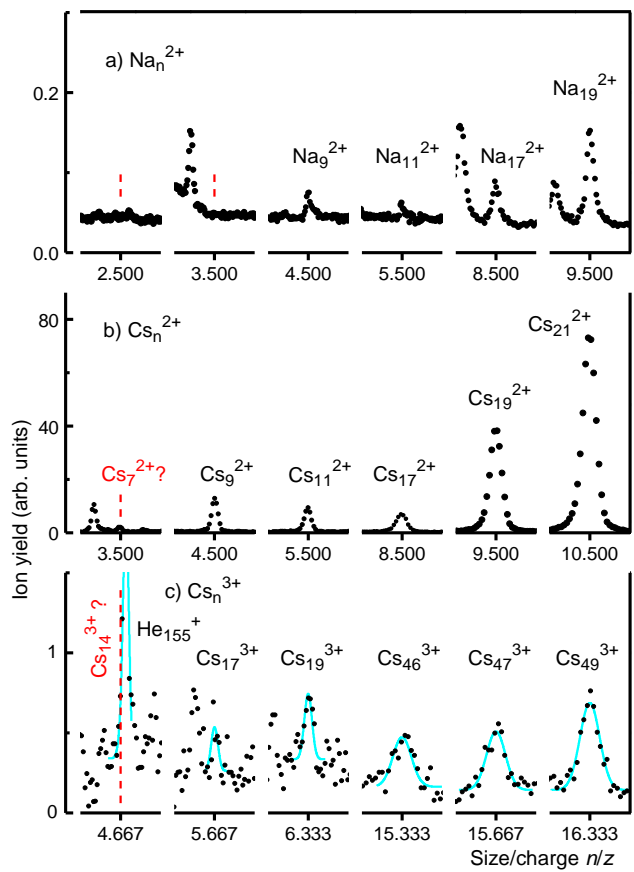

Fig. 1

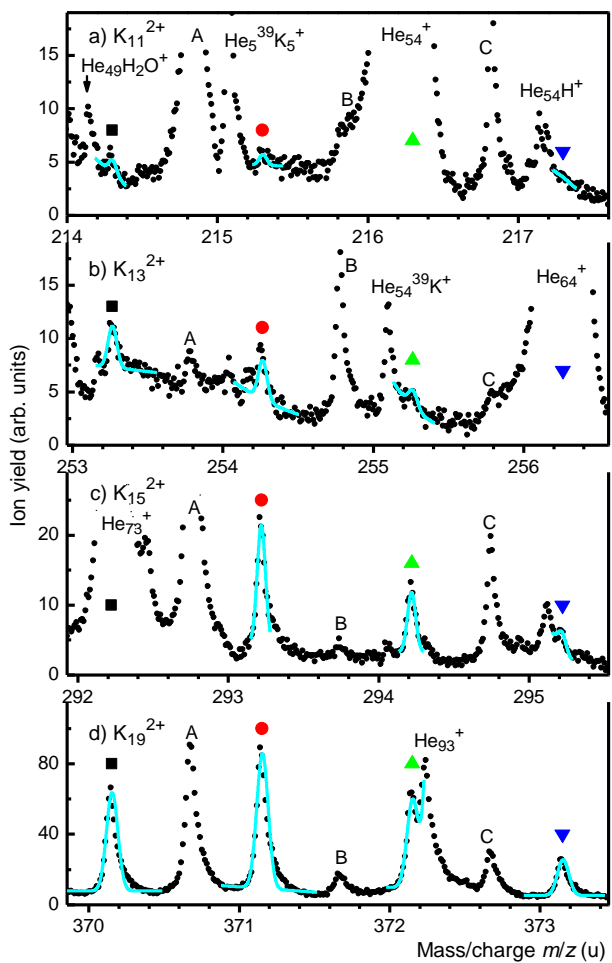

Fig. 2

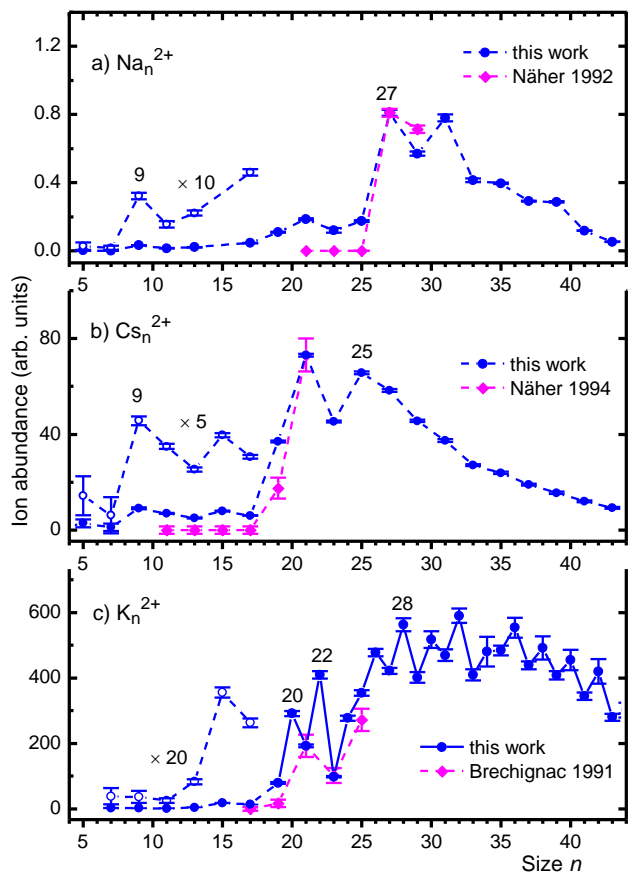

Fig. 3

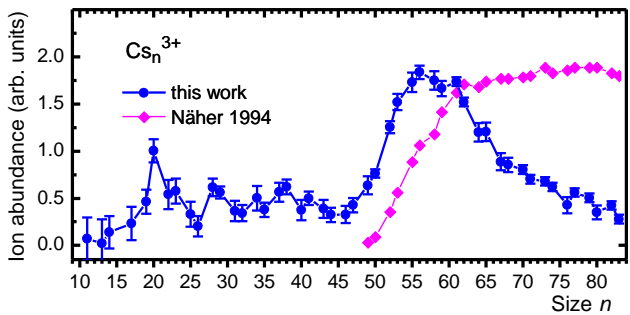

Fig. 4